# Numerical Study of Laminar and Turbulent Natural Convection from a Stack of Solid Horizontal Cylinders

**Subhasisa Rath**[*1], **Sukanta Kumar Dash**[2]

[1]School of Energy Science & Engineering, Indian Institute of Technology Kharagpur, 721302, India

[2]Department of Mechanical Engineering, Indian Institute of Technology Kharagpur, 721302, India

* Corresponding Author
E-mail: subhasisa.rath@gmail.com
Tel.: +91-9437255862
ORCID ID: https://orcid.org/0000-0002-4202-7434

**ABSTRACT**

Natural convection from a stack of isothermal solid horizontal cylinders has been investigated numerically in a three dimensional computational domain. Simulations were conducted in both laminar and turbulent flow regimes of Rayleigh number ($Ra$) spanning in the range ($10^4 \leq Ra \leq 10^8$) and ($10^{10} \leq Ra \leq 10^{13}$), respectively. In the present study, the length to diameter ratio of the cylinders has been varied in the range 0.5 to 20. Three different stack arrangements were considered for the numerical simulations by arranging three, six and ten number of cylinders in a triangular manner. The present computational study is able to appraise very interesting thermo-buoyant plume structures around the stack of cylinders. The average Nusselt number ($Nu$) shows a positive dependence on $Ra$ for all $L/D$. The average $Nu$ for a stack of three-cylinders is marginally higher than that of six-cylinders followed by ten-cylinders. Furthermore, at a particular $Ra$, $Nu$ is significantly higher for short cylinders (low $L/D$) and decreases with increase in $L/D$ up to 10 or 15 and remain constant for long cylinders. In addition, the present numerical results are also compared with the stack of hollow cylinders. A new Nusselt number correlation has been developed for different stacks as a function of $Ra$ and $L/D$, which would be useful to industrial practitioners and academic researchers.

## 1. Introduction

In many relevant physical situations, natural convection plays an indispensable role due to its passive nature and more significantly due to the absence of mechanical drive components. It makes the system more reliable, noise-free and reduces the extra power consumption to drive the flow. Natural convection heat transfer has numerous engineering applications in many industries as well as in real life practical situations. Natural convection cooling of heated horizontal cylinder or stack of heated cylinders has been a challenging topic in steel industries, where the present work is mostly intended for. Natural convection from horizontal solid cylinders has been extensively investigated by various researchers in the recent past. However, the available literature is almost nil for cylinders arranged in a stack and relatively less for multiple cylinders. A brief review on free convection from single and multiple solid horizontal cylinders is given here.

Natural convection heat transfer from a horizontal solid cylinder has been studied by Churchill and Chu [1] and their correlated expression for average *Nu* was found to be the most sited one. This correlation is the most referred one in the present days and also used for validation of the present numerical procedure. Three dimensional natural convection from an isothermal horizontal cylinder was investigated numerically by Acharya and Dash [2], which can be taken to be the most recent. The effect of *Ra* and *L/D* on heat transfer has been discussed in their work. For the entire range of *L/D* and *Ra,* the *Nu* for a single cylinder suspended in air was found to be higher than placed on the ground. Farouk and Guceri [3], numerically and theoretically studied laminar heat transfer from horizontal cylinders. The results were validated with the numerical results of Kuehn and Goldstein [4] and it was found that the *Nu* decreases along downstream of the cylinder. Farouk and Guceri [5] implemented the $(k - \varepsilon)$ turbulence model in the turbulent natural convection from a horizontal isothermal solid cylinder. The numerical results were compared with Churchill and Chu [1] and Kuehn & Goldstein [4]. Acharya and Dash [6] numerically investigated the free convection heat transfer from an external longitudinal finned hollow horizontal cylinder for Rayleigh number spanning in the laminar range of $10^4$ to $10^7$ by varying the parameters like L/D ratio of the cylinder, spacing between the fins and height of the fins. Kimura and Pop [7] theoretically and numerically studied the conjugate natural convection from a horizontal circular cylinder to demonstrate the effect of conduction in the core region on the natural convection from the cylinder surface. An approximate analytical solution has also been performed in their study. Laminar natural convection from a thick hollow horizontal cylinder placed on ground has been numerically investigated by Dash and Dash [8] by varying the length to outer diameter ratio (*L/D*) and inner to outer diameter ratio (*d/D*) in the range from 0.2–20 and 0.4–0.9, respectively. In their study, the average *Nu* was found to decrease with increase in thickness of the hollow cylinder at low L/D whereas, it was found to increase at higher *L/D*.

Research on multiple horizontal solid cylinders was started in 1991 by Paykoc et al. [9]. In their work, free convection from a pair of vertically placed horizontal solid cylinders was investigated theoretically and experimentally. Bejan et al. [10] analytically, numerically, and experimentally studied the laminar natural convection from an array of horizontal cylinders to investigate the optimal spacing. Computational study on free convection from a horizontal cylinder has been investigated by Cianfrini et al. [11] in laminar flow with the presence of another cylinder at downstream. The phenomena of natural convection from the staggered arrangement of twin horizontal cylinders were demonstrated experimentally and numerically by Heo et al. [12] in both laminar and turbulent flow regimes. Owing to preheating and side-flow effects on the thermal plume, the heat transfer rate of the upper cylinder were greatly affected by the presence of the bottom cylinder. Yuan et al. [13] numerically investigated free convection in a horizontal concentric annuli of different inner shapes to investigate the momentum and heat transfer characteristics including the thermal radiation.

Numerical investigation of free convection from a pair of two vertically placed isothermal cylinders was studied by Pelletier et al. [14] by considering water as the working medium. The heat loss from the top cylinder was found to be strongly influenced by the induced thermo-buoyant plume by the lower cylinder in their study. Natural convection from a vertical array of heated horizontal cylinders was numerically studied by Yu et al. [15] in molten salt nanofluids by varying the number of cylinders from 2 to 8 and pitches in the range 5 to 10. Interactive laminar free convection from two vertically separated isothermal horizontal cylinders has been numerically studied by Park and Chang [16] at $Pr$=0.7 and $Ra$ in the intermediate laminar range spanning from $10^4$ to $10^5$. The temperature imbalance ratio between the cylinders has also been presented for inter-cylinder spacing in the range of 2 to 4.

Liu et al. [17] numerically studied natural convection heat transfer from two attached horizontal cylinders (placed horizontally) in laminar range of $Ra$ spanning over 10 to $10^5$. The results were shown pictorially by streamlines and isotherms as well as the effect of Rayleigh number on local Nusselt number and local drag coefficients. Similar investigations on natural convection from two horizontal cylinders attached in a vertical plane has been done by Liu et al. [18] in an air medium. The results were compared with a single cylinder and it was found that owing to the interactions of their thermal plumes, the average Nusselt number for both the cylinders has been reduced compared to a single cylinder. Rath and Dash [19] numerically studied the two-dimensional natural convection from a stack of horizontal cylinders in both laminar and turbulent flow regimes which were valid for very long horizontal cylinders. Most recently, a three-dimensional numerical investigation on natural convection from a stack of thin hollow horizontal cylinders has been conducted by Rath and Dash [20]. Simulations were conducted in both laminar and turbulent flow regimes by varying $Ra$ and $L/D$ ratio of the cylinders to demonstrate the momentum and heat transfer characteristics from the inner surface of the cylinders. Whereas, the end effects for a stack of extremely thick hollow cylinders or stack of solid cylinders (i.e. steel rolls in steel industries) with varying $L/D$ ratio are not reported in the literature so far.

The present study is inspired in the direction of three dimensional natural convection cooling of solid cylinders arranged horizontally in a stack which are commonly seen in the bay of steel industries as shown in Fig. 1. This present study is an extension of the simulations proposed by Rath and Dash [19, 20] in order to investigate the end effect of different *L/D* ratios on both momentum and heat transfer characteristics. The natural convection heat transfer plays a significant role here for the cooling of such steel cylinders before that can be transported. Hence, the current study is intended to undertake the computational fluid dynamics (CFD) simulations for natural cooling of different stack arrangements of short or long solid horizontal cylinders. Numerical results are represented both qualitatively as well as quantitatively in terms of the overall heat loss and average Nusselt number for the stack, whereas the temperature and flow fields are pictorially visualized in terms of temperature contours and velocity vectors for different *L/D* and *Ra*.

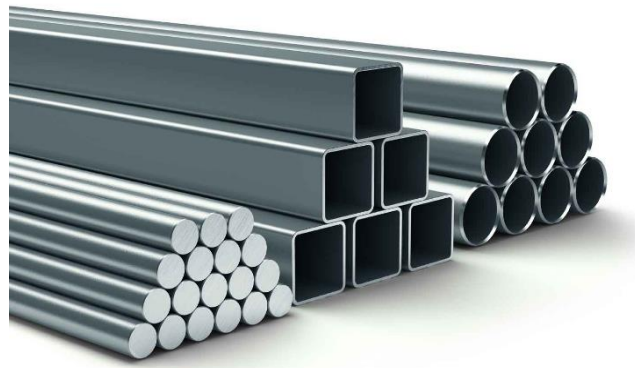

**Figure 1.** Stack arrangements of horizontal cylinders in steel industry bay for cooling [19]

## 2. Problem Description

The objective of the present work is to conduct a numerical study on natural convection from a stack of short or long horizontal cylinders. Hence, the numerical simulations are planned to perform in a *3D* computational domain, where cylinders of three, six or ten in numbers were arranged horizontally in a stack which are the commonly found stack structures in the bay.

Figure 2 represents a schematic diagram of the physical domain. Isothermal solid horizontal cylinders are placed over a flat adiabatic surface in a triangular manner to form a stack. A quiescent air medium of temperature '$T_\infty$' is surrounded over the stack. The lateral surface and the flat ends of the cylinders are maintained at isothermal conditions of uniform constant temperature ($T_w$), such that $T_w > T_\infty$.

The height (*H*), width (*W*) and length (*Z*) of the *3D* rectangular computational domain are considered in terms of the characteristics length scale ($L_C$). In the present study, the stack height is taken as the characteristic length ($L_C$). In the *3D* computational domain, the bottom surface is considered to be an adiabatic wall and the other five faces are treated as unconfined open boundaries. Gravity acts along the negative *y*-direction. For the present numerical simulations, length to diameter ratio (*L/D*) of the cylinders is taken in the range 0.5 to 20 and Rayleigh number (*Ra*) is considered in the range $10^4$ to $10^8$ for laminar and $10^{10}$ to $10^{13}$ for turbulent flow regimes.

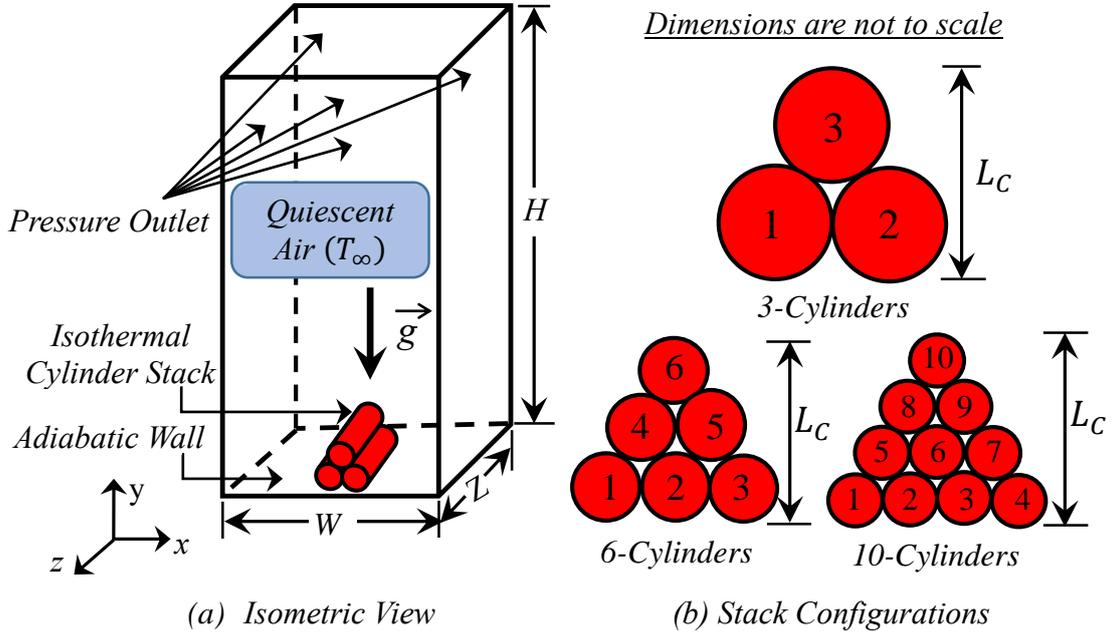

**Figure 2.** Schematic of the physical problem (a) Isometric view of the computational domain and (b) Different stack configurations

## *2.1. Mathematical Modelling*

### *2.1.1. Assumptions:*

In the present numerical study, the flow around the stack is assumed to be steady, incompressible, and buoyancy-driven. Radiation heat transfer and viscous dissipation are neglected in this study. Density is assumed to be constant in all the governing equations except in the body force term, where the temperature dependent density that drives the flow in natural convection is captured by taking Boussinesq approximation into account.

$$(\rho_\infty - \rho) \approx \rho\beta(T - T_\infty) \qquad (1)$$

Reynolds averaged Navier-Stokes (RANS) equations were employed for turbulent analysis and standard *k - ε* model was used to calculate the turbulent viscosity ($\mu_T$) in the present study.

### *2.1.2. Governing Equations:*

Based on the aforementioned assumptions, the governing conservation equations for the present numerical investigation are written in their tensorial form as follows:

*Continuity Equation:*

$$\frac{\partial}{\partial X_j}(\rho U_j) = 0 \tag{2}$$

*Momentum Equations:*

$$\frac{\partial}{\partial X_j}(\rho U_i U_j) = -\frac{\partial}{\partial X_i}\left[P + \frac{2\rho k}{3}\right] + \frac{\partial}{\partial X_j}\left[(\mu + \mu_T)\left(\frac{\partial U_i}{\partial X_j} + \frac{\partial U_j}{\partial X_i}\right)\right] + \rho g_i \beta (T - T_\infty)\delta_{i2} \tag{3}$$

*Thermal Energy Equation:*

$$\frac{\partial}{\partial X_j}(\rho U_j T) = \frac{\partial}{\partial X_j}\left[\left(\frac{\mu}{Pr} + \frac{\mu_T}{Pr_T}\right)\left(\frac{\partial T}{\partial X_j}\right)\right] \tag{4}$$

Turbulent eddy viscosity ($\mu_T$) cab be defined as:

$$\mu_T = \rho C_\mu \frac{k^2}{\varepsilon} \tag{5}$$

*Turbulent kinetic energy (k):*

$$\frac{\partial}{\partial X_j}(\rho k U_j) = \frac{\partial}{\partial X_j}\left[\left(\mu + \frac{\mu_T}{\sigma_K}\right)\frac{\partial k}{\partial X_j}\right] + G_K + G_b - \rho\varepsilon \tag{6}$$

*Turbulent dissipation rate ($\varepsilon$):*

$$\frac{\partial}{\partial X_j}(\rho \varepsilon U_j) = \frac{\partial}{\partial X_j}\left[\left(\mu + \frac{\mu_T}{\sigma_\varepsilon}\right)\frac{\partial \varepsilon}{\partial X_j}\right] + \frac{\varepsilon}{k}\left(C_{1\varepsilon} G_K + C_{1\varepsilon} C_{3\varepsilon} G_b - C_{2\varepsilon}\rho\varepsilon\right) \tag{7}$$

In equations (6) and (7), $G_K$ and $G_b$ represent the generation of turbulent kinetic energy due to mean velocity gradient and buoyancy effect, respectively.

$$G_K = -\mu_T \left(\frac{\partial U_i}{\partial X_j} + \frac{\partial U_j}{\partial X_i}\right)\left(\frac{\partial U_i}{\partial X_j}\right) \tag{8}$$

$$G_b = \beta g_j \frac{\mu_T}{Pr_T}\left(\frac{\partial T}{\partial X_j}\right) \tag{9}$$

The values of all the constants in equations (5), (6), and (7) are given below as per Launder and Spalding [21].

$$Pr_T = 0.85, \sigma_K = 1.0, \sigma_\varepsilon = 1.3, C_\mu = 0.09, C_{1\varepsilon} = 1.44, C_{2\varepsilon} = 1.92, C_{3\varepsilon} = tanh\left(\frac{U_x}{U_y}\right)$$

In the present study, for turbulent flow simulations, the $k$ and $\varepsilon$ equations (6) and (7) are solved along with the conservation governing equations (2-4), whereas for laminar flow simulations, the turbulent quantities are simply neglected by assigning a zero value to the turbulent viscosity ($\mu_T$). Since the objective is to compute an overall Nusselt number for the stack of solid cylinders in steady state, so the temperature solution within the solid cylinder may not be solved for. Hence, we have not done that here rather by specifying the surface temperature to be constant one can arrive at a flow and temperature field which can provide the Nusselt number for the stack.

### 2.1.3. Near Wall Function:

According to the law of the wall, the mean velocity of turbulent flow at a certain point is proportional to the log of the distance between the wall and that point.

$$U^* = \frac{1}{k} ln(Ey^*) \tag{10}$$

$$U^* = \frac{U_P C_\mu^{(1/4)} k_P^{(1/2)}}{\tau_W / \rho} \tag{11}$$

$$y^* = \frac{\rho C_\mu^{(1/4)} k_P^{(1/2)} y_P}{\mu} \tag{12}$$

Where, $E$ = empirical constant = 9.793, $k$ = Von-Karman constant = 0.4187, $U_P$ = mean velocity of the fluid at point $P$, $C_\mu$ = empirical constant = 0.09, $y_p$ = wall distance from point $P$, $k_p$ = turbulence kinetic energy.

In the present study of turbulent flow analysis, the "*Scalable wall function*" model has been implemented for interlinking the variables of $k$ and $\varepsilon$ to the wall surface which is found to be the most suitable one. The limiting value of $y^*$ in "*Scalable wall function*" model is 11.225 and calculated as:

$$\tilde{y}^* = \text{Max}(y^*, y^*_{limit}) \tag{13}$$

### 2.1.4. Boundary Conditions:

The boundary conditions which are imposed to the computational domain as shown in Figure 2 (for solving the governing equations) are summarized below with their mathematical expression.

*On the cylinder surfaces*: The lateral surface along with the end surfaces of the cylinders are assigned as wall boundaries with no slip and no penetration conditions in addition to isothermal boundary condition.

$$U_x = U_y = U_z = 0 \text{ and } T = T_w \tag{14}$$

*The bottom surface of the domain*: The adiabatic wall boundary condition along with no slip and no penetration conditions are imposed on the bottom flat surface over which the cylinders are placed by considering the earth surface as a non-conducting surface.

$$\dot{q}_w = 0 \text{ and } U_x = U_y = U_z = 0 \tag{15}$$

*At the outer boundaries*: Pressure outlet boundary condition is imposed on all the five outer faces of the computational domain (four side faces and the top face).

$$P = P_{atm} \text{ (zero gauge) and } T = T_\infty \text{ (backflow temperature)} \quad (16)$$

*At symmetry*: A reflective symmetry condition is imposed on the longitudinal cross-sectional plane of the stack as shown in Figure 3.

$$U_x = U_z = 0, \quad \frac{\partial U_y}{\partial y} = 0 \text{ and } \frac{\partial T}{\partial y} = 0 \quad (17)$$

## 2.2. Geometrical Parameters & Thermo-physical Properties

In the present study, constant thermo-physical properties of air are considered at the mean temperature ($T_m$). All the geometrical parameters and fluid properties which are implemented in the present numerical simulations are given in Table 1.

**Table 1.** Geometrical parameters and fluid properties

| Properties | Value | Parameters | Value |
|---|---|---|---|
| $\rho_\infty$ | 1.1765 kg/m$^3$ | $L_C$ | 1.4535 m |
| $C_P$ | 1007.6 J/kg-K | $D_{(3\text{-Cylinders})}$ | 0.7789 m |
| $K$ | 0.028 W/m-K | $D_{(6\text{-Cylinders})}$ | 0.532 m |
| $\mu$ | $1.97 \times 10^{-5}$ kg/m-s | $D_{(10\text{-Cylinders})}$ | 0.404 m |
| $\beta$ | 0.003077 K$^{-1}$ | | |

## 2.3. Heat Transfer Parameters

The total heat loss from the entire stack of cylinders is due to the convection heat transfer and can be calculated as:

$$Q = hA_t(T_w - T_\infty) \quad (18)$$

Where $A_t$ is the total surface area of the stack

$A_t = A_S + A_C$
$A_S$ = Total curved surface area; $A_S = n\,(\pi DL)$
$A_C$ = Total end cross-sectional area; $A_C = 2n\{(\pi/4) \times D^2\}$

The total heat transfer rate can be evaluated by integrating the temperature field over the entire stack surface.

$$Q = -\int K \frac{\partial T}{\partial \hat{n}} dA \quad (19)$$

Where $\hat{n}$ is the outward normal to the cylinder surface

The average heat transfer coefficient (h) is thus computed:

$$h = \frac{Q}{A_t(T_w - T_\infty)} \tag{20}$$

The average surface Nusselt number is calculated as:

$$Nu = \frac{hL_c}{K} = \frac{QL_c}{A_t(T_w - T_\infty)K} \tag{21}$$

Rayleigh number based on the characteristic length scale is formulated as:

$$Ra = \frac{g\beta(T_w - T_\infty)L^3}{\vartheta\alpha} \tag{22}$$

## *2.4. Numerical Procedures*

The governing conservation equations were solved by using the imposed boundary conditions in a finite-volume based Computational Fluid Dynamics (CFD) solver *ANSYS-Fluent-15*. Steady state with pressure based solver was used for the present work. For the turbulent analysis, the standard $k$-$\varepsilon$ model was employed for calculating the turbulent eddy viscosity ($\mu_T$) with full buoyancy effect. Scalable wall function was implemented as near wall treatment for the turbulent model. In the present study, for coupling the pressure-velocity terms, SIMPLE algorithm was employed which was found to be the most stable one. Body force weighted scheme has been implemented for the discretization of pressure. For the discretization of the convective terms in momentum and energy equations along with the turbulent kinetic energy and turbulent dissipation rate, initially, First Order Upwind scheme was used for a trial convergence and for the more accurate result we shifted over to Second Order after the initial convergence was obtained. For a smooth convergence, the convergence criteria for energy equation was set to $10^{-6}$ and for all other equations, it was set to $10^{-4}$. The under-relaxation factors used for convergence in the present simulation are shown in Table.2. After getting a converged solution, both domain independence and grid independence tests were carried out for the final results.

**Table 2.** Under-relaxation factors used in the simulation

| Pressure | Density | Body Force | Momentum | Turbulent Kinetic Energy | Turbulent Dissipation Rate | Turbulent Viscosity | Energy |
|---|---|---|---|---|---|---|---|
| 0.3 | 1 | 1 | 0.4 | 0.8 | 0.8 | 1 | 1 |

## *2.5. Grid Distributions*

For the numerical study, in order to solve the governing equations, it is necessary to discretize the computational domain into a number of grid points. Figure 3 shows the grid distribution in the domain. In the present numerical study, a symmetry of the flow can be anticipated from the geometry about the longitudinal plane of the stack. Hence, a plane-symmetry boundary condition has been imposed for numerical simulations to reduce the computational resources. Near the cylinder walls, the gradient of field properties would vary drastically. Hence, very fine cells are provided near the cylinder walls by unstructured tetrahedral cells in order to capture the velocity and temperature distributions adjacent to the wall whereas, relatively coarser mesh is provided far away from the stack by structured hexahedral cells as shown in Figure 3. For solving in turbulent regime, the first cell height ($y*$) has been maintained in the log-law region by using scalable wall function ($y* >11.225$).

## 2.6. Domain Independence Test

In order to make sure that the numerical results remain independent of the computational size of the unconfined domain, a domain-independent study was carried out in the present study for the extreme case of long cylinders ($L/D=20$) in laminar regime of $Ra=10^8$. At the same time, this domain independence test also keeps the requisite numerical resources at an optimal level. The length of the computational domain was fixed at ($Z=L+4L_C$), which would be more than sufficient for the present study according to our past experience. The width and height of the domain were varied to different $L_C$ as per Table 3. From Table 3 it can be concluded that the width and the height can be fixed to ($W=4\times L_C$) and ($H=12\times L_C$), respectively since the intended quantity ($Nu$) does not change substantially (within 1%) after this.

**Table 3.** Effect of domain size on average Nusselt number for a stack of three-cylinders at $L/D=20$ and $Ra=10^8$

| Domain Size: ($W\times H$) | $2L_C \times 4L_C$ | $3L_C \times 8L_C$ | $4L_C \times 12L_C$ | $5L_C \times 16L_C$ |
|---|---|---|---|---|
| Nusselt Number: ($Nu$) | 35.92 | 38.17 | 39.35 | 39.67 |

## 2.7. Grid Independence Test

A Grid independence test was conducted to ensure that the computations remain independent of the grid size. Figure 4 shows the average surface Nu variation with the level of refinements of the grids for the stack of three-cylinders with $L/D=1$ and $Ra=10^6$. Initially, the solution was obtained with coarse grids (*Level 0*) and subsequent refinement was given near the cylinder surfaces until the variation in Nusselt number has been found to be negligible. From Figure 4, it can be seen that the *Level-3* refinement with 562563 cells was found sufficient for this particular case of $L/D=1$, where the variation in *Nu* with *Level-4* refinement was less than 1%. Hence, *Level-3* refinement was used for simulations of all the cases of $L/D=1$, since a further refinement would increase the computation time and space without much improvement in accuracy of results. For all other types of stacks and $L/D$, a similar grid independence study was carried out before reporting any result.

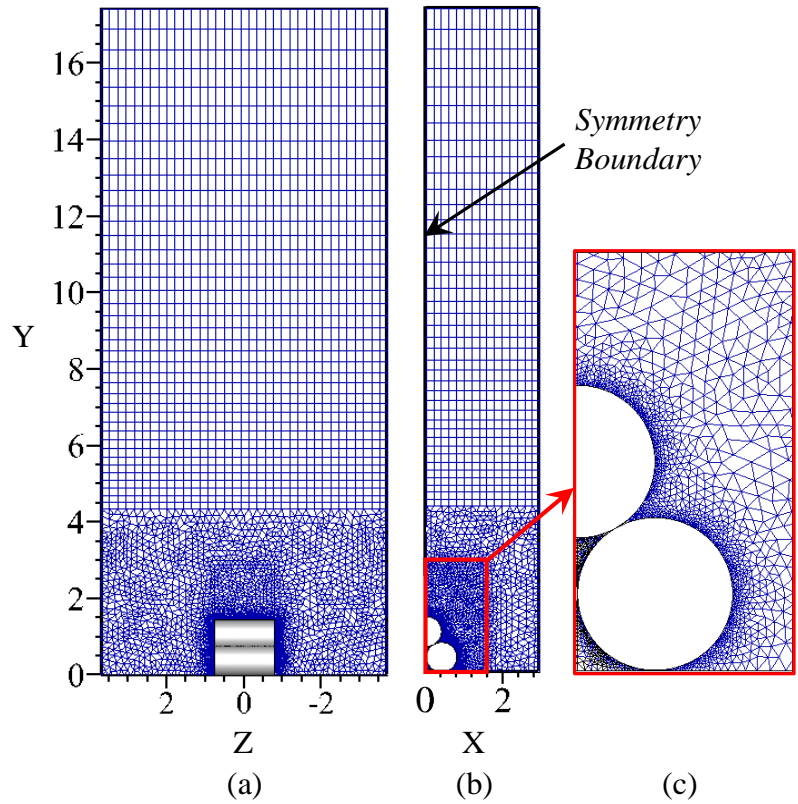

**Figure 3.** Arrangement of computational grids in the domain, (a) Transverse section view, (b) Cross-sectional view and (c) Blown up view near the cylinders

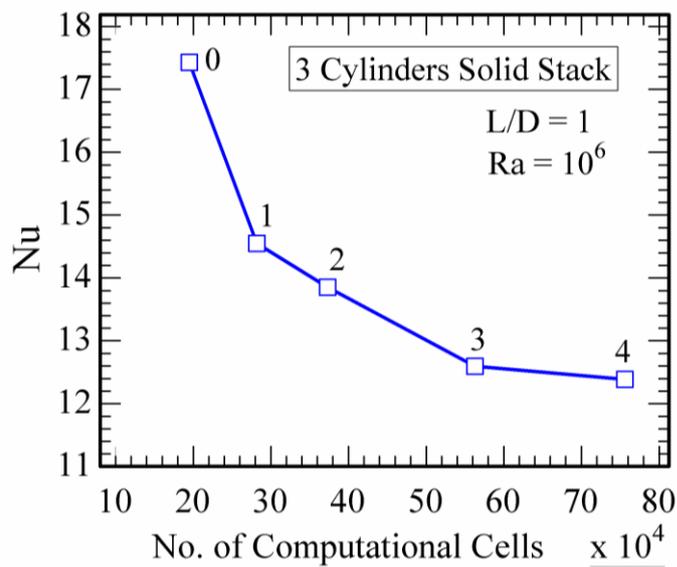

**Figure 4.** Variation of *Nu* with level of refinements of the computational cells

## 3. Results and Discussion

### 3.1. Model Validation

To check the accuracy and suitability of the numerical methodology used in the present computations, we have validated with prior existing experimental and numerical results for a *2D* single solid horizontal cylinder suspended in air by performing some special numerical simulations. Figure 5(a) shows a comparison of average *Nu* between the present simulation results and experimental correlations of Churchill and Chu [1], Morgan [22], and McAdams [23] along with the experimental data of Chung et al. [24] at high *Ra*. It can be seen that our present computations shows a very good agreement with the experimental data and correlations in the laminar range. In order to check the accuracy in local results, we have further compared our results with the numerical results of Saitosh [25], Wang et al. [26], and Kuhen and Goldstein [4] by manifesting the variation of local *Nu* over the surface of the horizontal cylinder with circumferential angle as shown in Figure 5(b). The special simulation result for a single solid cylinder in air matches fairly well with the preceding reported literature with a maximum of 2% deviation. Hence, we have gained enough confidence in our present numerical methodology which would predict much accurate results for natural convection.

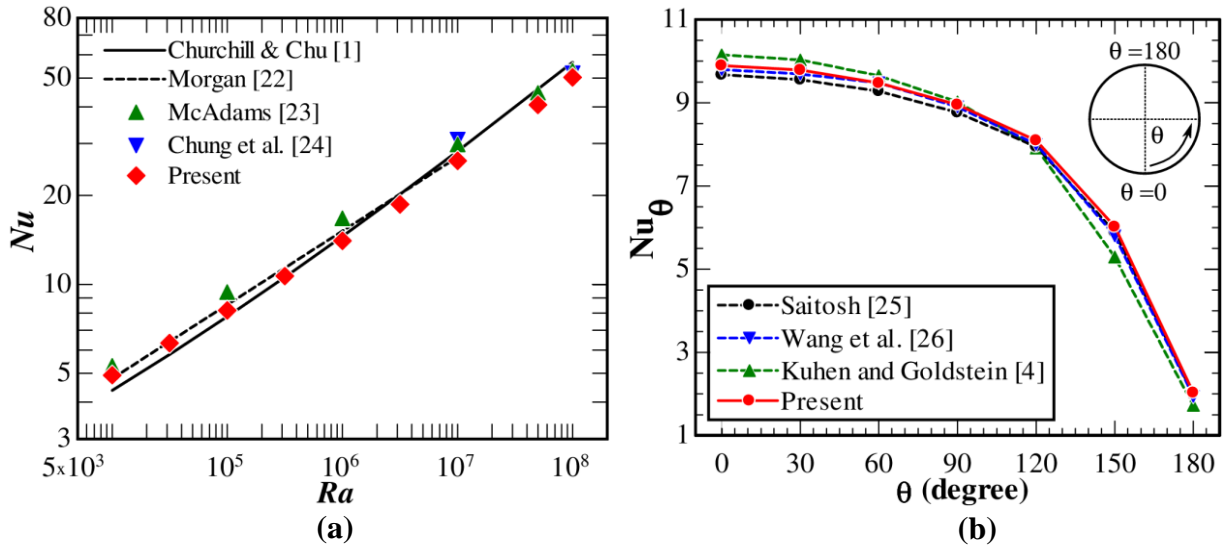

**Figure 5**. Comparison with previous results; (a) variation of average *Nu* as a function of *Ra* and (b) variation of local *Nu* over the cylinder as a function of angle at $Ra=10^5$

### 3.2. Effect on average Nusselt number

Figure 6 shows the variations of average surface *Nu* with *L/D* of the solid cylinders at different *Ra* in both laminar and turbulent flow regimes. It can be seen that with an increase in *L/D*, there is a progressive decrease in *Nu* and beyond *L/D* of 15, *Nu* becomes almost constant in laminar regime whereas, Nu marginally decreases in turbulent regimes (can be seen from log scale). Since the edge effect of the flow plays a prominent role at low *L/D* so *Nu* at low *L/D* is found to be higher than that at high *L/D*. With increase in *Ra*, the average surface *Nu* increases for all the cases in the present study due to the buoyancy effect becoming stronger with increased *Ra*. When the flow is turbulent, the average *Nu* is much higher compared to that at laminar flow. Figure 7 shows a comparison between different stacks in both laminar and turbulent regimes. In both laminar as well as turbulent flows, the average *Nu* for the stack of three-cylinders is seen to be moderately higher than the stack of six-cylinders followed by ten-cylinders due to more exposed surface area to ambient air for the stack of three-cylinders resulting in higher heat transfer coefficient compared to the stack of six and ten-cylinders.

### *3.3. Effect on total heat loss*

The variation of heat loss with *L/D* can be seen in Figure 8 for different *Ra*. The total heat loss is found to increase with an increase in *L/D* of the solid cylinders due to the increase in heat transfer surface area of the stacks. It can also be found that as *Ra* increases, the total convective heat transfer rate increases exponentially in both laminar as well as in turbulent regimes. When *Ra* increases, the strength of the buoyancy effect becomes stronger which helps to remove more heat from the surface of the cylinders. The buoyancy plume is predicted to be stronger in turbulent regime compared to laminar regime which can be seen from the thermal plume structure in Figure 11 and Figure 12. Although the characteristic length scales of all the three type of stacks are same, the diameter of the cylinders for different stacks are considered to be different in order to fix this characteristic length scale (can be found from Table 1). Hence, for a particular *L/D*, the length of the cylinders in the stack of three-cylinders is quite longer than the stack of six-cylinders followed by the stack of ten-cylinders. For this reason, it can be seen from Figure 8 that with an increase in *L/D*, heat loss from the stack of three-cylinders is marginally higher compared to the stack of six or ten-cylinders.

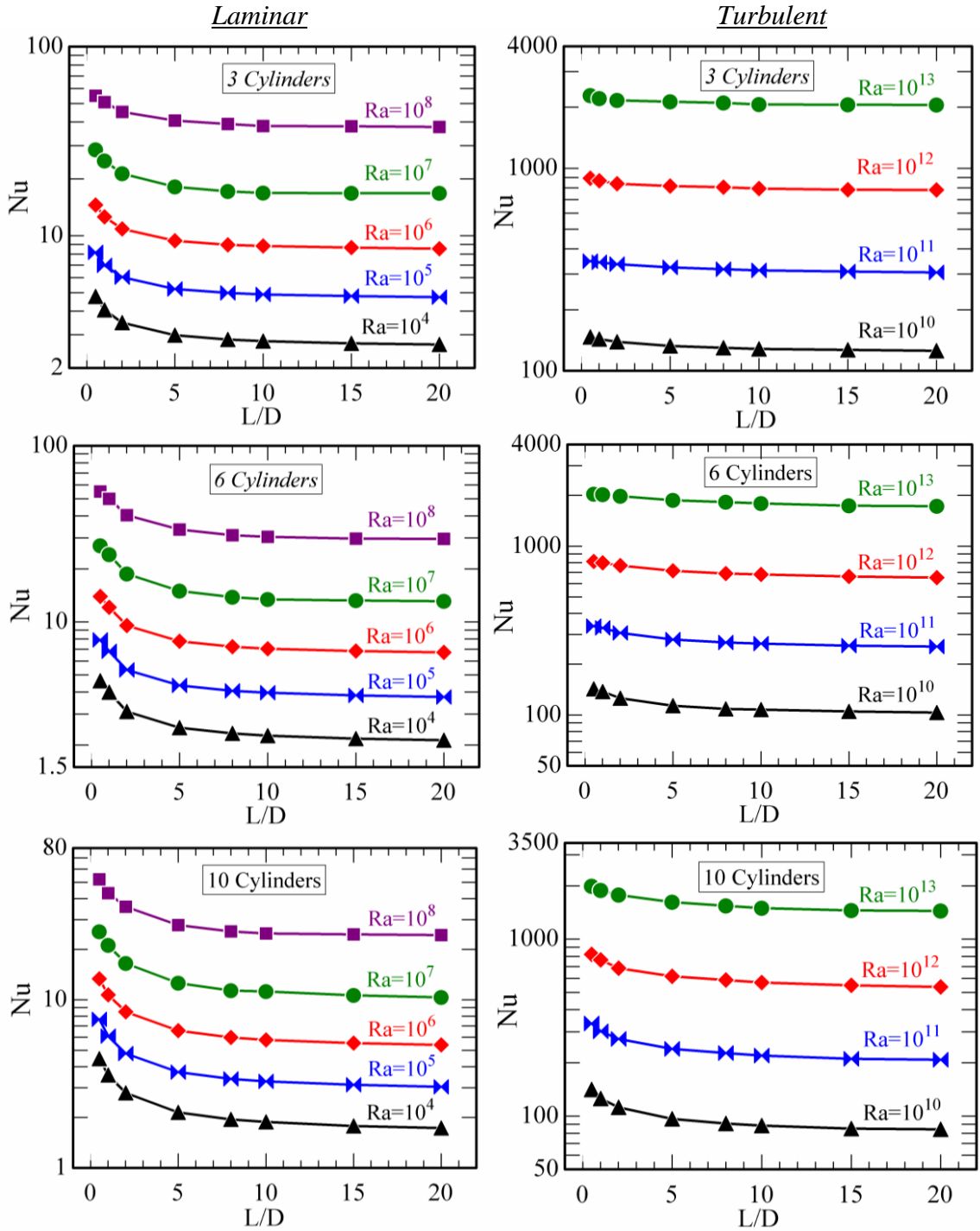

**Figure 6.** Dependence of *Nu* as a function of *L/D* and *Ra* for laminar and turbulent flows

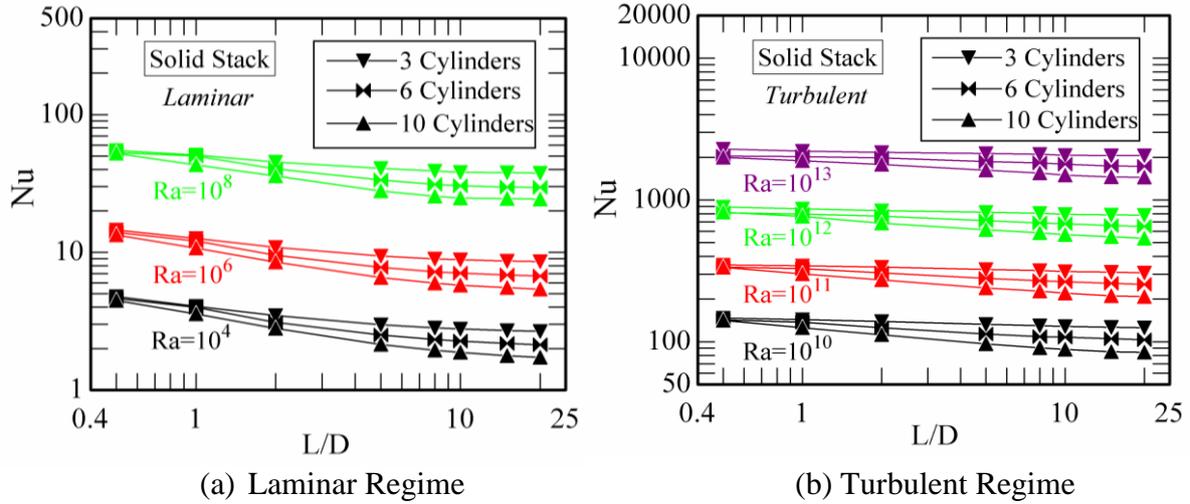

(a) Laminar Regime  (b) Turbulent Regime

**Figure 7.** Dependence of Nu as a function of *L/D* and *Ra*, a comparison between different stacks

*3.4. Contribution of cylinder end loss to total heat loss*

Figure 9 shows the ratio of heat loss from the flat ends of the solid cylinders to the total heat loss of a stack of three-cylinders for different *Ra* with different *L/D*, to quantify the contribution of cylinder ends on heat loss. It has been found that at low *L/D* (starting from 0.5) the contribution of cylinder ends to heat transfer is significantly higher and gradually decreases with an increase in *L/D*. In laminar flow regime up to $Ra=10^6$, the contribution of flat-ends to heat loss increases for low *L/D* up to 2 and remains constant for high *L/D* (beyond 2) whereas, for *Ra* beyond $10^6$ this contribution gradually decreases for all *L/D*. Furthermore, heat transfer from the side flat end of the cylinders would remain to be high since the flat end would get air flow from the longitudinal direction as well as from the radial direction whereas the curved surface would get air flow from the radial direction only. So when *L/D* is low the heat loss from the ends would predominate over the heat loss from the curved faces. As *L/D* increases the curved surface area increases whereas the flat end surface area remains fixed. So, the predominance of heat loss from the flat ends is lost now when compared with the curved surface heat loss because the curved surface area has increased a lot due to the increase in *L/D*. Hence, the ratio of end heat loss to that of the total loss diminishes progressively with increase in *L/D*. The heat loss in turbulent flow is extremely high, so the end loss does not predominate over the curved surface heat loss. Hence, the ratio of end heat loss to total heat loss always falls with *Ra* which was not the case in laminar flow.

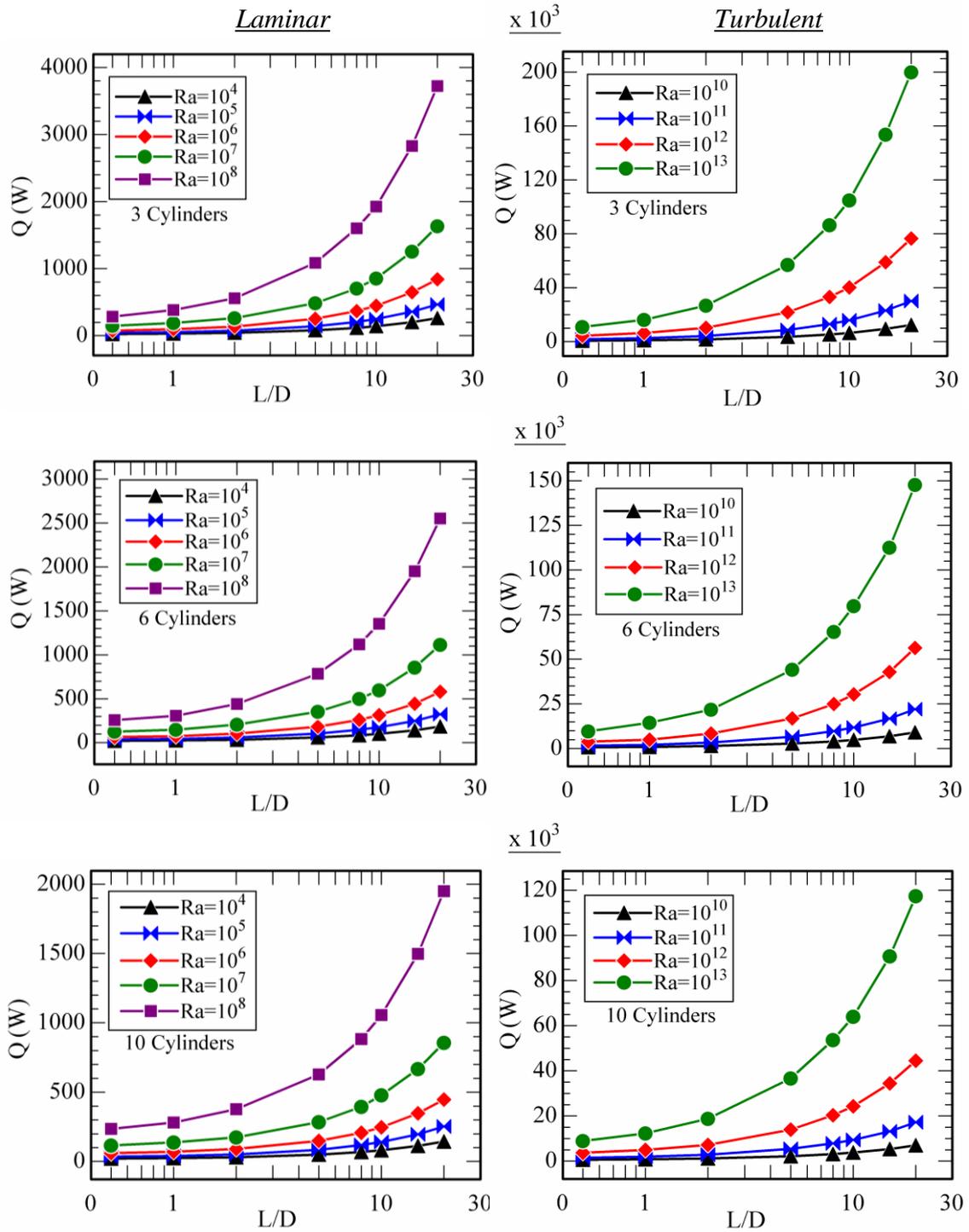

**Figure 8.** Variation of heat loss as a function of *L/D* for different *Ra*, in both laminar and turbulent regimes

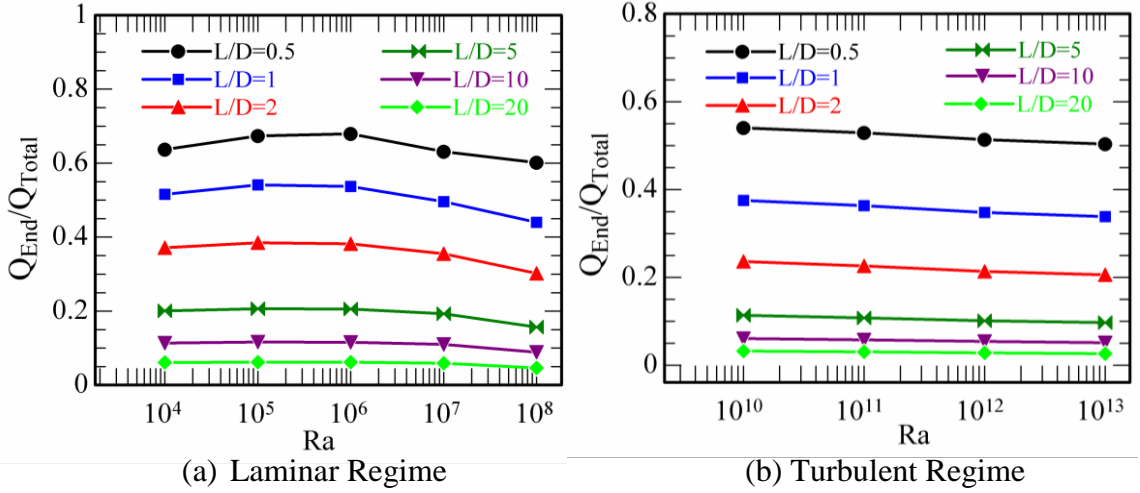

(a) Laminar Regime        (b) Turbulent Regime

**Figure 9**. Contribution of cylinder ends to total heat loss as a function of *Ra* for different *L/D* of a stack of three-cylinders

## *3.5. A comparison with 2D results*

In the present work, an attempt was made to justify the *2D* numerical simulations of Rath and Dash [19] by comparing the extreme case of *L/D*=20 of the *3D* simulations. For this comparison, a case of three-cylinders in the stack was considered for the whole range of *Ra* considered in this study. As *2D* model signifies the *L/D* of the cylinders to be very large so we plan to choose the extreme case of *L/D*=20 of the present *3D* simulation results. Figure 10 shows the variation of average Nusselt number with *Ra* for both *2D* and *3D* simulation results. It has been found that starting from *Ra*=$10^4$ to $10^7$, the *2D* result shows a very good agreement with the *3D* results with a very negligible error (within 6%) which signifies that *L/D* of 20 is more than sufficient to justify the *2D* model. For *Ra* beyond $10^7$, the *2D* result shows a marginal error (more than 14%) and this error increases with increase in *Ra* which signifies that *L/D* of 20 is not sufficient to justify the *2D* model. So it can be concluded that for low *Ra*, the *2D* modeling is justified for *L/D* more than 20 and this justification limit of *L/D* increases with increase in *Ra* beyond $10^7$. Hence, it is suggested to consider *L/D* > 25 or 30 to justify a *2D* result in turbulent regime.

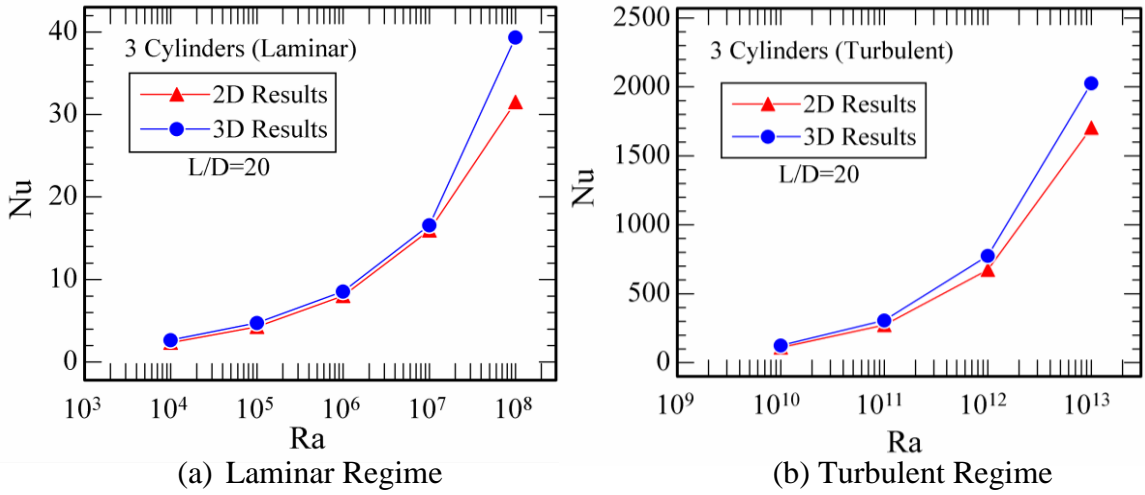

| (a) Laminar Regime | (b) Turbulent Regime |

**Figure 10**. Comparison of *Nu* between *2D* and *3D* results for a stack of three-cylinders

*3.6. A Comparison with a Stack of hollow Cylinders:*

In the present study, an attempt has also been made to compare the present results (stack of solid cylinders) with the numerical results of Rath and Dash [20] for a stack of hollow cylinders. Figure 11 shows the variation of *Nu* as a function of *L/D* for a stack of six solid and hollow cylinder in both laminar and turbulent flows. It can be marked that up to a *Ra* of $10^6$ the *Nu* for the stack of solid cylinders is marginally higher compared to the stack of hollow cylinders for all *L/D*. With increase in *L/D* this difference between the Nu for the stack of hollow and solid cylinders gradually diminishes and beyond $Ra=10^7$ for low *L/D* up to 8, the *Nu* for the stack of hollow cylinders is more compared to that of the stack of solid cylinders in both laminar as well as in turbulent flows. As the stacks are placed on ground, the stack of hollow cylinders would lose heat from its inner surfaces, outer surfaces and from the portion of the inter-cylinder gaps more effectively due to the penetration of buoyant plume into the hollow cylinders stack whereas, the stack of solid cylinders cannot lose heat more effectively from its bottom surfaces. Hence, the *Nu* for the stack of hollow cylinders is found to be higher at high *Ra* and low *L/D* (*L/D* < 5) compared to the stack of solid cylinders. But for (*L/D* > 5), the average *Nu* of the stack of hollow cylinders goes down and becomes less than that of the solid cylinders stack. For the solid cylinders, heat will be lost from the flat ends whereas for the hollow cylinders this does not happen and moreover, with high *L/D,* the heat loss from the inner surface of the hollow cylinder falls drastically compared to the case of low *L/D*. So, when *L/D* increases beyond 5 the heat loss becomes more for the solid stack compared to that for the hollow stack of cylinders and hence, the *Nu* becomes more for the stack of solid cylinders compared to that of hollow cylinders.

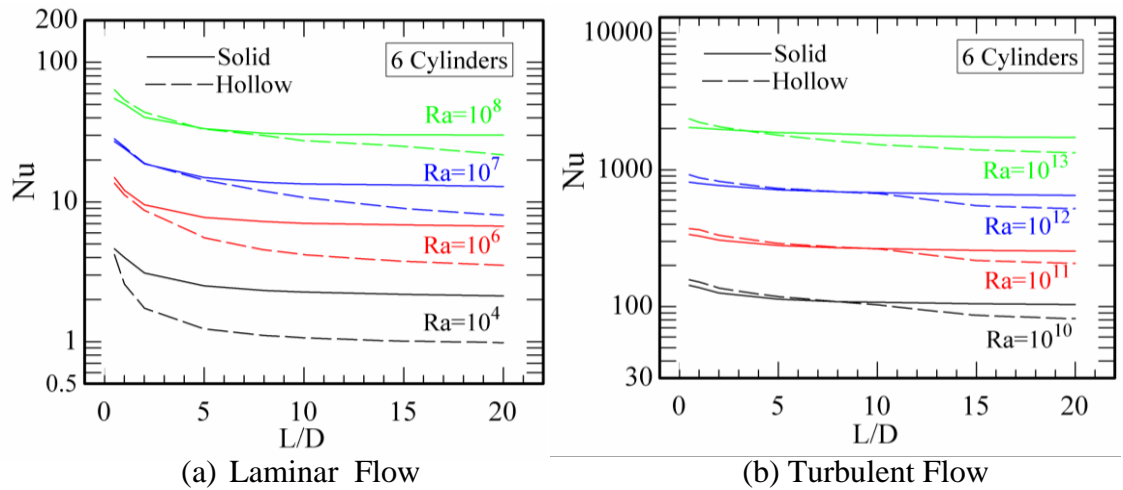

(a) Laminar Flow      (b) Turbulent Flow

**Figure 11.** Comparison of *Nu* between stack of solid and hollow cylinders

### *3.7. Effect of Ra on Thermal Plume*

The contours of static temperature around the stack of solid cylinders are shown in Figure 12 as a function of *Ra* in the laminar range ($10^4 \leq Ra \leq 10^8$). It also shows a comparative analysis of the thermal plume in different cross-sectional planes of the stack which has never been seen in any of the literature so far. In Figure 12, column 1 and 2 shows the transverse cross-sectional plane at the end and middle of the cylinders, respectively. Whereas, column 3 shows the longitudinal plane in the middle of the stack. At low *Ra* (*Ra*=$10^4$), the thermo-buoyant plume is seen to be quite thick around the stack and when *Ra* increases, owing to the stronger buoyancy-driven flow over the stacks, the thermal boundary layer gradually becoming thinner. It can also be noticed that at low *Ra*, heat removal from the internal gaps of the cylinder stack is poor since the temperature plume looks to be thicker. Whereas, when *Ra* increases, the gaps between the cylinders try to cool down gradually (heat removal becomes better) from the flat-end towards the inside of the stack. The thermal plume at the end plane shows higher temperature above the stack compared to the plume at the mid-cross-sectional plane. The end planes of the cylinders receive air flow from radial as well as from the longitudinal direction and hence more heat lost to the air so the air temperature above the stack remains higher compared to the mid-cross-sectional plane where heat is lost to ambient air coming only from the radial direction. These thermo-buoyant plume structures at different cross-sectional planes look to be very interesting and many physical insights can be drawn from such pictorial views.

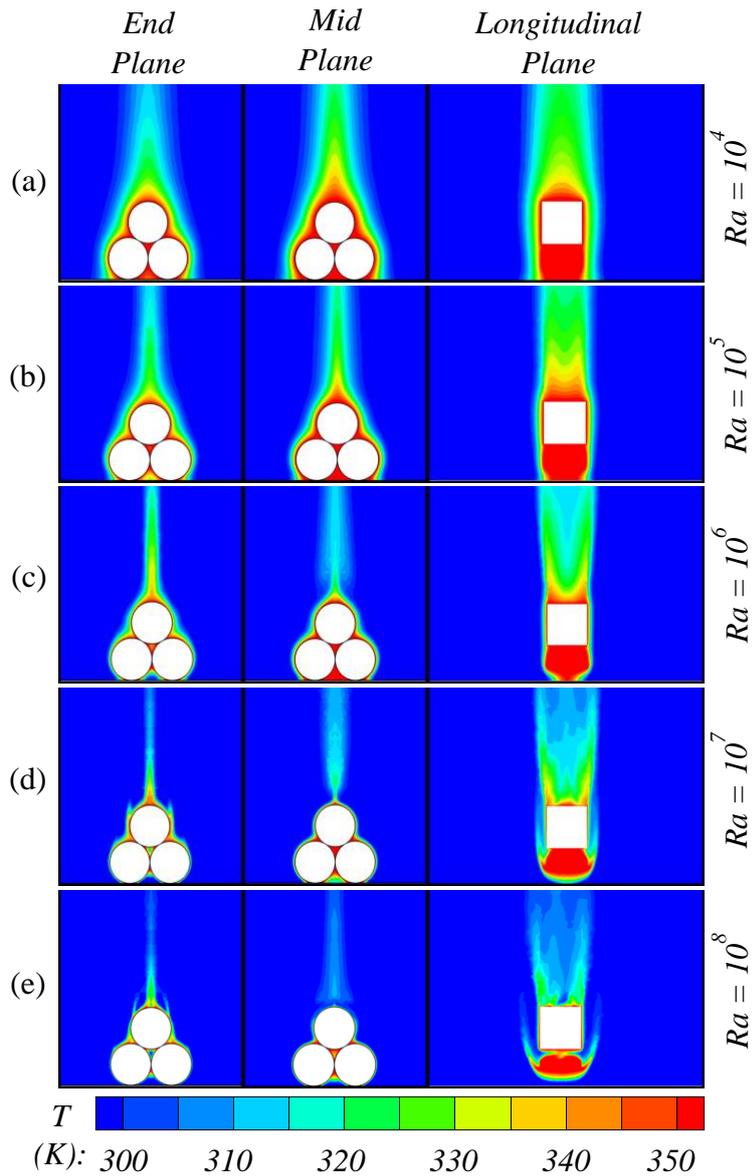

**Figure 12**. Contours of temperature as a function of *Ra* in laminar regime, for a stack of three-cylinders having *L/D*=1

Figure 13 shows the temperature contours for all the three types of stacks as a function of *Ra* in turbulent regime. It can be noticed that at low *Ra* of $10^{10}$, the temperature plume is thick compared to high *Ra* in turbulent regime but it is significantly thin compared to any high *Ra* of the laminar regime. The thicker temperature plume at low *Ra* results in less heat transfer to ambient air compared to a high *Ra*. In addition, the core of the thermal plume (above the stack) is seen to be much cooler at high *Ra*. Hence, *Nu* is found to be comparatively low in laminar flow compared to that of turbulent flow.

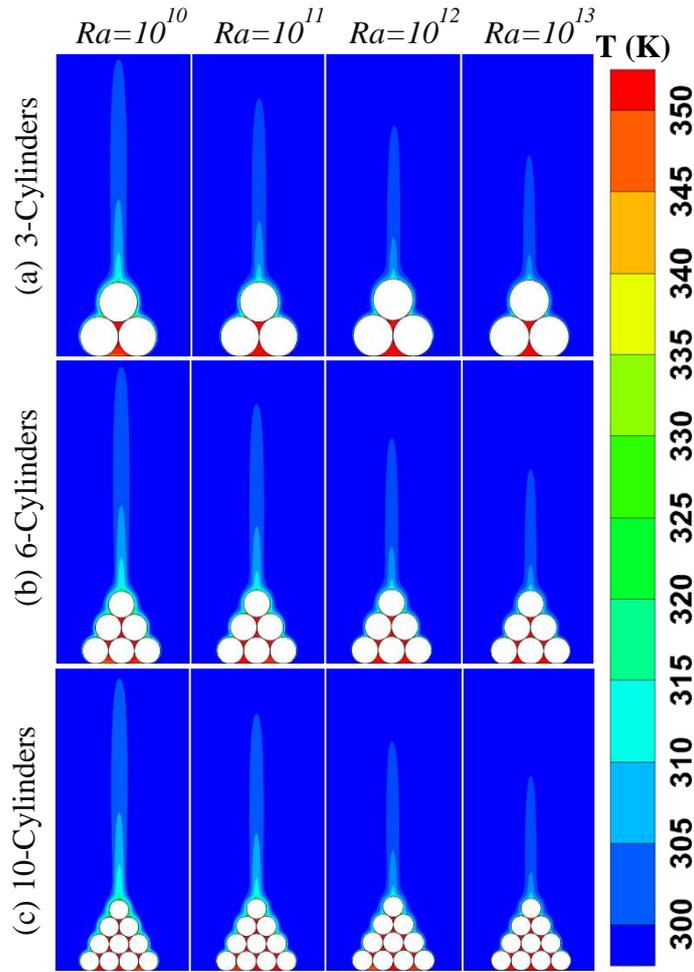

**Figure 13.** Contours of static temperature as a function of *Ra* in turbulent regime at mid transverse cross-sectional plane

*3.8. Effect of L/D on Thermal Plume*

The effect of *L/D* of the solid cylinders on thermal plume is shown in Figure 14 at $Ra=10^6$ for a stack of three cylinders. It can be seen from the transverse cross-sections that the thermal boundary layer around the stack is very thin at the end of the cylinders and gradually becoming thicker towards inside mid-section of the cylinders for low *L/D* of 0.5. The thermal plume just above the stack is quite hot at the end cross-sectional plane compared to the mid-cross-sectional plane. This is because of the edge effect of the solid cylinders due to which more heat is lost to the ambient air causing the thermal plume to be hotter above the stack. It can also be seen from Figure 14 that on the mid-cross-sectional plane, for low *L/D* (starting from *L/D*=0.5) the thermal plume is quite thick at the top of the stack and gradually becoming thin when *L/D* increases, indicating more heat is removed as the surface area increases. At low *L/D*, the surface of the cylinders in the gaps of the stack can lose heat due to the penetration of buoyant plume into the inside portion of the gaps. As *L/D* increases, the heat loss from the cylinder surfaces in the gaps of the stack falls because the buoyant plume is not able to penetrate into a longer distance through the gap. The pictorial visualization of these thermo-buoyant plume structures for different stacks over the entire range of *Ra* with different *L/D* is quite interesting from an academic viewpoint because many practical insights of flow can be visualized from such pictures.

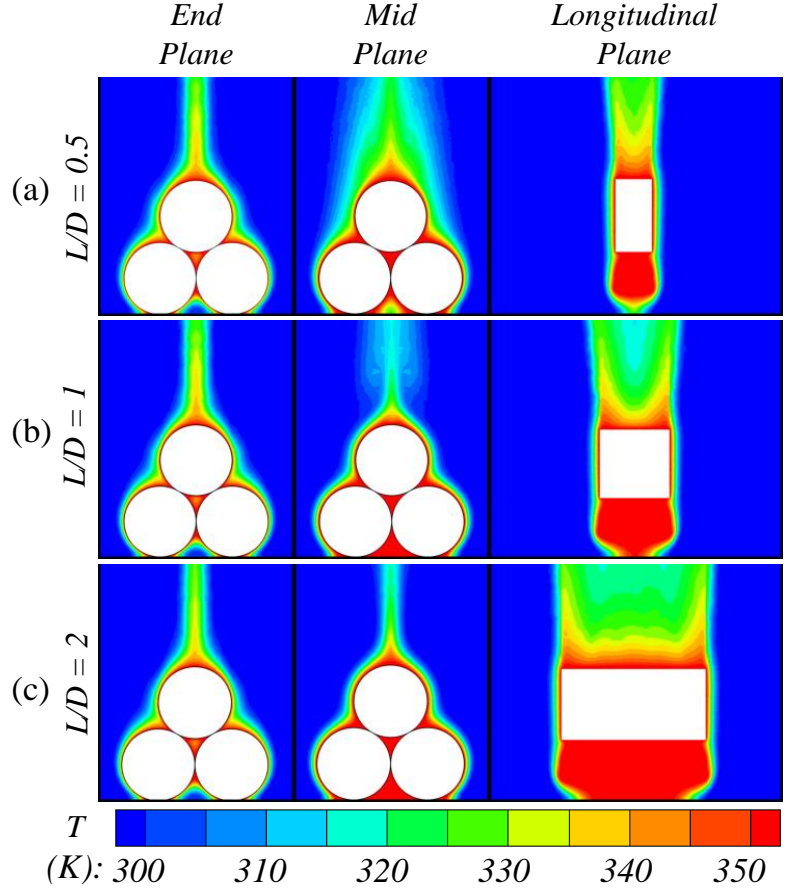

**Figure 14.** Temperature contours with varying *L/D*, at *Ra*=$10^6$

## *3.9. Effect of L/D on Flow Field*

The velocity vectors on the longitudinal cross-sectional plane of a stack of three-cylinders are shown in Figure 15 for varying *L/D* at *Ra*=$10^4$. As the bottom surface of the domain is adiabatic, it can be seen that the surroundings quiescent air is drawn towards the stack only from the side faces of the domain and the plume rises up towards the top of the domain due to the buoyancy effect by taking heat from the heated solid cylinders. Due to the no-slip wall boundary condition, the tangential velocity is found to be zero over the cylinder surfaces and the growth of the velocity boundary layer can be seen when we move away from the cylinder surfaces. The magnitude of the velocity can be seen to be higher in the core of the buoyant plume for all the cases in this present study. With an increase in *L/D*, the thickness of the flow plume increases along the longitudinal plane. It can also be seen from Figure 15 that for a particular *Ra*, the velocity boundary layer at the flat end of the cylinder is almost same for different *L/D*. It is very interesting to see the flow plume penetrating into the gaps of the cylinders in the stack and coming out through the gaps. As predicted from the temperature contours in Figure 14, for low *L/D*, the flow plume is able to reach to the mid-section of the stack through the gaps and removes heat from the heated surfaces of the cylinders. Whereas for high *L/D*, it is not able to reach to the mid-section, it only takes heat from very small lengths in the gaps from the flat end faces.

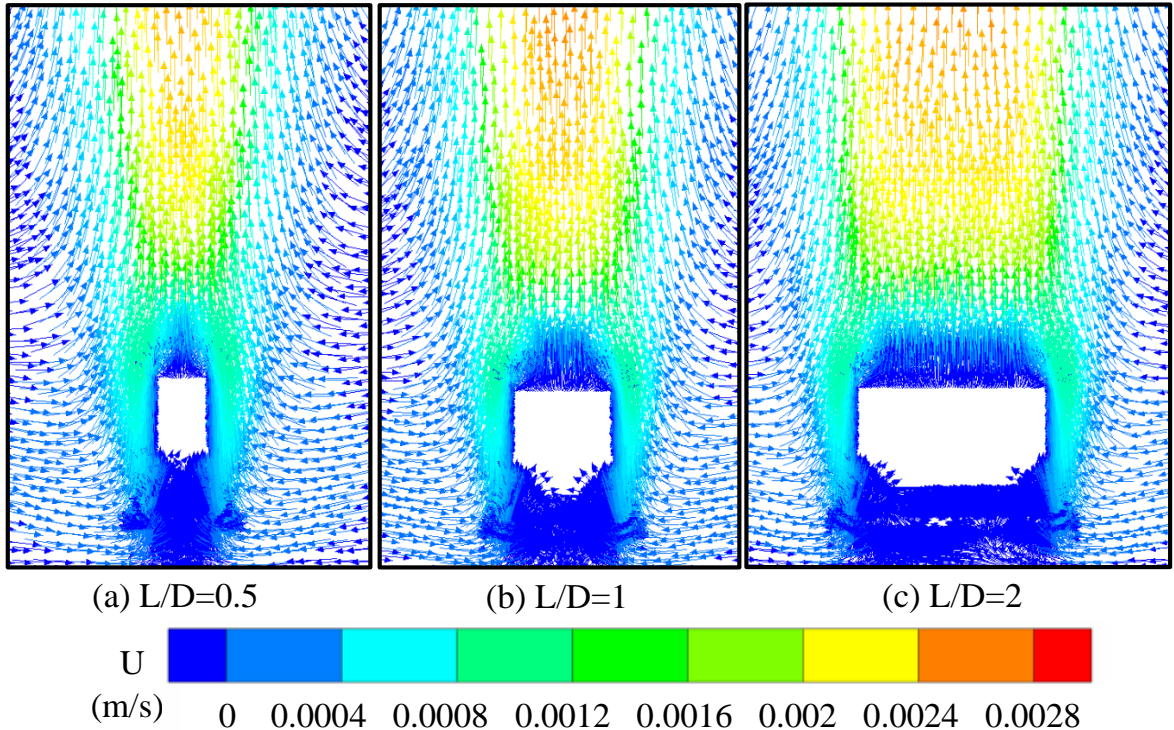

**Figure 15.** Velocity vectors with varying *L/D* along the longitudinal plane at $Ra=10^4$

## *3.10. Effect of Ra on Flow Field*

Figure 16 shows the velocity vector around the stack of three-cylinders as a function of *Ra* for *L/D*=2. At low Ra, more fluid levitates from the bottom and flows vertically upward by sliding over the stack of cylinders whereas, at high *Ra*, the strength of the buoyancy increases and more fluid draws from both sides of the stack. In addition, at high *Ra*, the plume velocity near the stack surface is found to be higher compared to that of low *Ra*, which can be seen from the velocity scale attached to each picture. It can be marked that the velocity boundary layer thickness is quite thick at low *Ra* of $10^4$ and gradually becoming thinner around the cylinders when *Ra* increases, which can happen in both laminar as well as in turbulent regimes. It can also be noticed that at high *Ra* the rising flow glides fast over the cylinder surface and can't roll down the curved surface of the cylinder well due to its high velocity, whereas at low *Ra* the attachment of the flow plume and the rolling of the plume over the curved surface occurs very smoothly. From Figure 16 it can be concluded that the velocity of flow increases with increase in *Ra* resulting in more heat loss from the cylinders, which was already predicted from the temperature contours discussed in section 3.6.

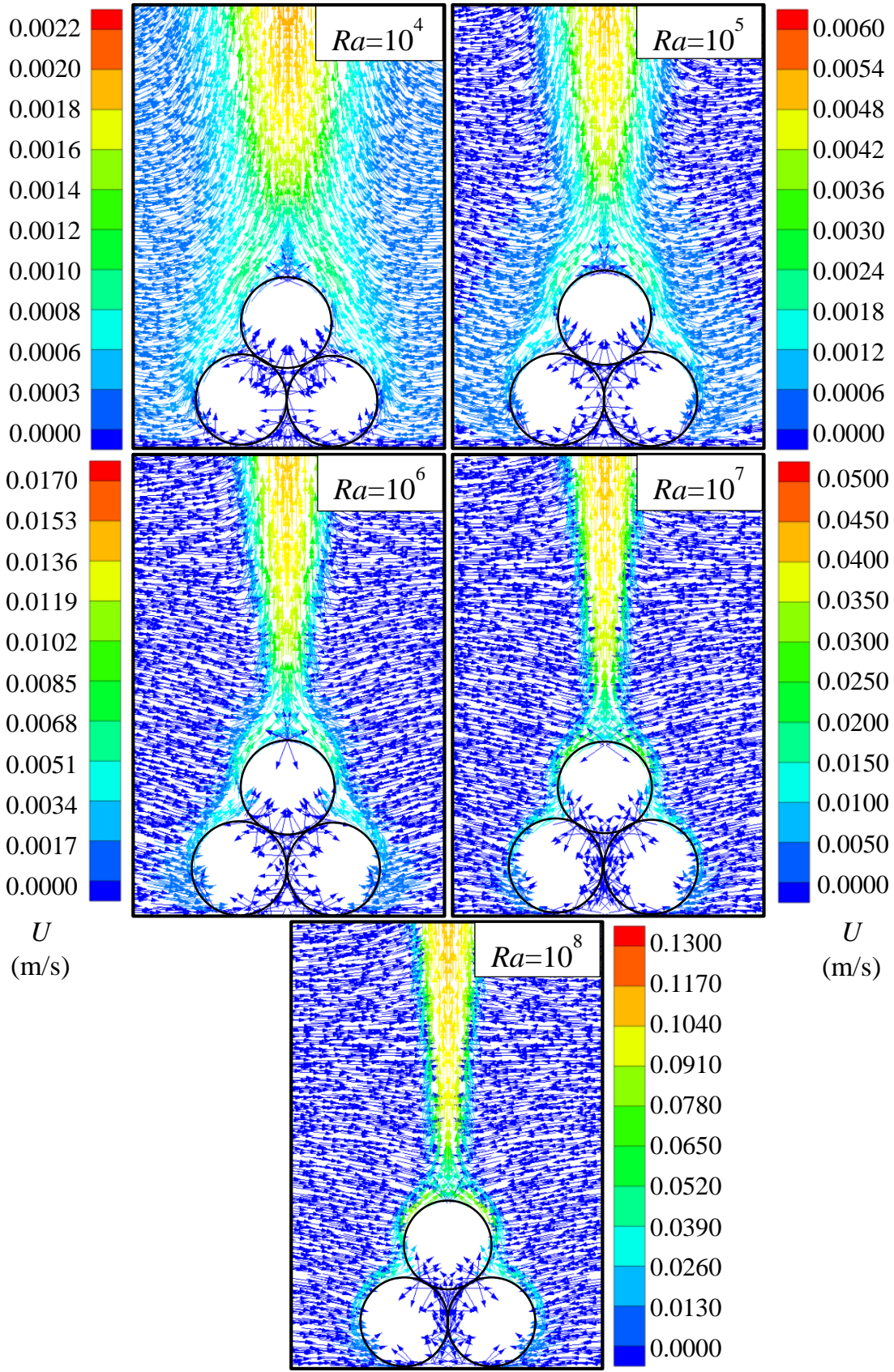

**Figure 16.** Velocity vector as a function of *R*a for *L/D*=2 on the Mid-cross-sectional plane

## 4. Nusselt Number Correlation

An empirical correlation has been developed in the present study for the average *Nu* by performing a non-linear regression analysis of the numerical results, which would be useful to the engineers for industrial calculations. In the present numerical analysis, as *Nu* is a function of *Ra* and *L/D* so the functional relationship can be expressed as:

$$Nu = f(Ra, L/D) \quad (23)$$

The correlating equation (24) has been developed for as a function of *Ra* and *L/D* valid for *Ra* in the range of ($10^4 \leq Ra \leq 10^8$) in laminar and ($10^{10} \leq Ra \leq 10^{13}$) in turbulent flow, for *L/D* in the range 0.5 to 20.

$$Nu = C_0 + C_1(Ra)^{n_1} + C_2(L/D)^{n_2} + C_3(Ra)^{n_3}(L/D)^{n_4} \quad (24)$$

Where, $C_0, C_1, C_2, C_3$ and $n_1, n_2, n_3, n_4$ are the correlation constants given in Table 4. Figure 17 shows the computed and the predicted values of *Nu* for a stack of ten-cylinders in both laminar and turbulent flow regimes. A clear agreement of all the data points is seen within the accuracy range of ($\pm 8\%$) where the coefficient of correlation ($R^2$) was found to be above 0.99 for all the cases. Similar plots were also obtained for the stacks of three and six-cylinders which are not shown here for the sake of brevity.

**Table 4.** Correlation constants for the stacks of three, six, and ten-cylinders

| *Regime* | *Stack* | *Correlation Constants* |
|---|---|---|
| Laminar | Three-Cylinders | $C_0 = 0.4803, C_1 = 0.0495, C_2 = 1.8106, C_3 = 0.0535$<br>$n_1 = 0.3555, n_2 = -0.4397, n_3 = 0.2963, n_4 = -0.5676$ |
| | Six-Cylinders | $C_0 = 6.3657, C_1 = 0.0199, C_2 = -4.4681, C_3 = 0.0935$<br>$n_1 = 0.3843, n_2 = 0.0590, n_3 = 0.2953, n_4 = -0.5140$ |
| | Ten-Cylinders | $C_0 = 4.3574, C_1 = 0.0162, C_2 = -2.6509, C_3 = 0.0889$<br>$n_1 = 0.3831, n_2 = 0.0958, n_3 = 0.2982, n_4 = -0.5720$ |
| Turbulent | Three-Cylinders | $C_0 = 1851.4, C_1 = 85.0060, C_2 = -25.8539, C_3 = -343.062$<br>$n_1 = 0.1738, n_2 = -1.0762, n_3 = 0.1265, n_4 = 0.0028$ |
| | Six-Cylinders | $C_0 = -1784.822, C_1 = 9230.772, C_2 = -416.2628, C_3 = 1.3754$<br>$n_1 = -0.0703, n_2 = -0.0160, n_3 = 0.2576, n_4 = -0.0341$ |
| | Ten-Cylinders | $C_0 = 3811.299, C_1 = -834.1704, C_2 = -990.6942, C_3 = 89.6587$<br>$n_1 = 0.0863, n_2 = -0.0719, n_3 = 0.1577, n_4 = -0.0228$ |

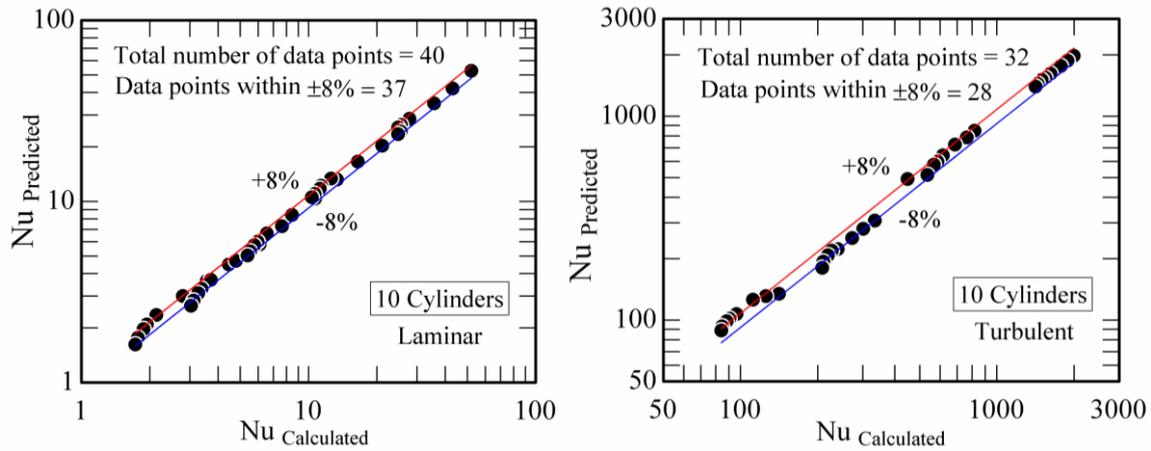

**Figure 17.** Computed and predicted values of *Nu* for a stack of ten-cylinders

## 5. Conclusions

In this work, a numerical study has been performed to investigate the cooling of a stack of heated horizontal solid cylinders by natural convection heat transfer for a wide range of *L/D* ($0.5 \leq L/D \leq 20$) and Rayleigh number in the range of ($10^4 \leq Ra \leq 10^8$) and ($10^{10} \leq Ra \leq 10^{13}$) for laminar and turbulent flow regimes, respectively. The governing conservation equations were solved numerically in a *3D* computational domain to predict the momentum and heat transfer characteristic around the stack in terms of the effect of *Ra* and *L/D* on *Nu*, heat loss, flow field, and thermal plume. From the present computational analysis, the important observations are enumerated below.

(a) The average *Nu* was found to increase with an increase in *Ra* for all the cases of *L/D* in both laminar and turbulent regime. Whereas, with an increase in *L/D*, *Nu* decreases and remains almost constant beyond *L/D*=15.

(b) For all type of stacks, the total heat loss was found to increase significantly with increase in *Ra* and *L/D*, owing to increase in the strength of buoyancy and heat transfer surface area, respectively.

(c) For the entire range of *Ra* and *L/D* considered in the present study, the average Nu and the total heat transfer were found to be higher for the stack of three-cylinders compared to the stack of six-cylinders followed by ten-cylinders.

(d) Pictorial visualization of the thermo-buoyant plume structures and flow field through velocity vectors over the stack for different cases are elucidated in the present study, which would provide a better understanding of heat transfer and flow physics to the researchers.

(e) A comparison with *2D* results was also conducted to justify the *2D* simulations for different *Ra*. For *Ra* up to $10^7$, *2D* simulations can be done for a stack of cylinders provided the *L/D* is more than 20 (average *Nu* remains within 6% when *2D* and *3D* results are compared). For *Ra* greater than $10^7$, *2D* simulations in laminar or turbulent flow can be done provided the *L/D* is more than 25 or 30.

**Nomenclature**

- $A_t$     total heat transfer area, $m^2$
- $A_S$     total curved surface area, $m^2$
- $A_C$     total end cross-sectional area, $m^2$
- $D$     diameter of the cylinder, $m$
- $L$     length of the cylinders, $m$
- $n$     number of cylinders in the stack
- $L_c$     characteristic length scale, $m$
- $g$     acceleration due to gravity, $m/s^2$
- $K$     thermal conductivity of air, $W/m\text{-}K$
- $T$     temperature, $K$
- $P$     pressure, $N/m^2$
- $Q$     heat transfer rate, $W$
- $h$     heat transfer coefficient, $W/m^2\text{-}K$
- $Nu$     average Nusselt number
- $\dot{q}_w$     wall heat flux, $W/m^2$;
- $Pr$     Prandtl number
- $Ra$     Rayleigh number;
- $k$     turbulent kinetic energy, $m^2/s^2$
- $U$     flow velocity, $m/s$
- $U^*$     dimensionless velocity
- $y^*$     dimensionless wall unit
- $X$     Cartesian coordinate
- $i, j$     index notations

**Greek Symbols**

- $\alpha$     thermal diffusivity, $m^2/s$
- $\beta$     thermal expansion coefficient, $K^{-1}$
- $\mu$     dynamic viscosity, $kg/m\text{-}s$
- $\mu_T$     turbulent viscosity or eddy viscosity, $m^2/s$
- $\vartheta$     kinematic viscosity, $m^2/s$
- $\varepsilon$     turbulent dissipation rate, $m^2/s^3$
- $\rho$     density of the fluid, $kg/m^3$
- $\theta$     circumferential angle, degree
- $\delta_{i2}$     Kronecker delta; $\delta_{i2} = \begin{cases} 1; & \text{if } i \equiv 2 \\ 0; & \text{if } i \neq 2 \end{cases}$

**Subscripts**

- $x,y,z$     x-direction, y-direction, z-direction
- $w$     wall
- $m$     mean
- $\infty$     quiescent air
- $atm$     atmospheric
- $T$     turbulent
- $\theta$     local

**Acknowledgment**

The present research work was carried out in the department of mechanical engineering at the Indian Institute of Technology Kharagpur, India.

This research did not receive any specific grant from funding agencies in the public, commercial, or not-for-profit sectors.